\documentclass[10pt]{article}

\usepackage{amsmath}
\usepackage{amssymb}

\usepackage{graphicx}

\usepackage{cite}

\usepackage{color} 

\usepackage{url}

\usepackage[labelfont=bf,labelsep=period,justification=raggedright]{caption}

\makeatletter
\renewcommand{\@biblabel}[1]{\quad#1.}
\makeatother

\date{}

\begin{document}

% Title must be 150 characters or less
\begin{flushleft}
{\Large
\textbf{Cooperation and contagion in web-based, networked public goods experiments}
}
% Insert Author names, affiliations and corresponding author email.
\\
Siddharth Suri and Duncan J. Watts 
\\
\bf Microeconomics and Social Systems, Yahoo! Research, New York, NY, USA
\\
E-mail: suri@yahoo-inc.com, djw@yahoo-inc.com
\end{flushleft}

% Please keep the abstract between 250 and 300 words
\section*{Abstract}
A longstanding idea in the literature on human cooperation is that
cooperation should be reinforced when conditional cooperators are more
likely to interact.  In the context of social networks, this idea implies
that cooperation should fare better in highly clustered networks such as
cliques than in networks with low clustering such as random networks.  To
test this hypothesis, we conducted a series of web-based experiments, in
which 24 individuals played a local public goods game arranged on one of
five network topologies that varied between disconnected cliques and a
random regular graph.  In contrast with previous theoretical work, we
found that network topology had no significant effect on average
contributions. This result implies either that individuals are not conditional
cooperators, or else that cooperation does not benefit from positive
reinforcement between connected neighbors.  We then tested both of these
possibilities in two subsequent series of experiments in which artificial
“seed” players were introduced, making either full or zero contributions.
First, we found that although players did generally behave like conditional
cooperators, they were as likely to decrease their contributions in response to 
low contributing neighbors as they were to increase their contributions in
response to high contributing neighbors. 
Second, we found that positive effects of
cooperation were contagious only to direct neighbors in the network.
In total we report on 113 human subjects experiments, highlighting the
speed, flexibility, and cost-effectiveness of web-based experiments over those conducted in
physical labs.

% Please keep the Author Summary between 150 and 200 words
% Use first person. PLoS ONE authors please skip this step. 
% Author Summary not valid for PLoS ONE submissions.   
% \section*{Author Summary}

\section*{Introduction}
Why, and under what conditions, presumptively selfish individuals cooperate
is a prevailing question in social science that has stimulated an
extraordinary range of explanations, many of which have focused on the
strategic benefits of cooperation. For example, although displays of
altruism may appear to run counter to an individual's self-interest, it is
possible to show that if one assumes that individuals possess sufficiently
strong other-regarding preferences, then altruism may in fact convey
selfish benefits as well~\cite{fehr-fairness}. Moreover in a social
context, behavior that appears purely altruistic may also accrue individual
benefits either because others explicitly reward pro-social
behavior~\cite{rand-positive, ule-punishment} or punish selfish
behavior~\cite{bowles-reciprocity, fehr-altruistic, boyd-punishment,
  gurerk-advantage}. Finally, individuals may be rewarded indirectly for
cooperating, either because a good reputation conveys other transactional
benefits~\cite{milinski-reputation}, or because altruistic behavior can be
viewed as a signal of reproductive fitness~\cite{gintis-signalling}.

In addition to explanations that focus on individual strategies, a
longstanding idea is that cooperative behavior might arise as a consequence
of the population structure itself~\cite{hamilton-genetical}. Although
initially proposed in the context of evolutionary biology, this idea has
particular relevance for social dilemmas among human actors, where the
total population is large, but the effects of any one individual's actions
fall disproportionately on a relatively small set of neighbors determined
either by spatial or social proximity.  For example, smog or acid rain
causing pollutants disproportionately impact geographically proximate
populations; thus one can think of the game as playing out on some
approximation of a spatial lattice.  Correspondingly, the benefit derived
from social networking sites (e.g.\ Facebook) is highly dependent on the
activities and contributions of one's immediate social acquaintances, whose
identities in turn depend some complicated mixture of social and spatial
distance~\cite{watts-identity}.  Because in either case an individual's
neighbors are themselves connected to others, who are in turn connected to
others still, and so on, the dynamics of social dilemmas can be thought of
as taking place on extended networks~\cite{newman-survey,
  strogatz-exploring}.  In these settings, outcomes of interest, such as
aggregate levels of cooperation, plausibly depend on the structure of the
network as well as on the strategies of the individuals in the
population~\cite{nowak-structured}.

There are two main reasons to suspect that cooperation should depend on
network structure.  The first reason is that many theoretical models of
social dilemmas assume that cooperation is conditional, in the sense that
an individual will only cooperate on the condition that its partners are
also cooperating. Arguably the clearest example of the principle of
conditional cooperation is the celebrated Tit-For-Tat strategy, which has
consistently been shown to outperform more exploitative strategies in a
range of simulation studies, in large part because it performs well when
interacting with other cooperative
strategies~\cite{axelrod-evolution-book}. In addition, related strategies
have also been proposed that generalize the idea of conditional cooperation
to multi-player settings~\cite{watts-thesis, glance-outbreak}, usually by
specifying some form of threshold requirement---i.e.\ ``I will cooperate if
at least X of my neighbors cooperated last round, else I will defect.''
Regardless of the specifics of the rule, the implication of these results
for networks is that networks characterized by high levels of local
clustering~\cite{watts-small-world}, meaning that an individual's neighbors
are also likely to be neighbors of each other, ought to sustain higher
aggregate levels of cooperation than populations in which individuals are
randomly mixed~\cite{axelrod-evolution}. Put another way, local
reinforcement would imply that when an individual's neighbors also interact
with each other, they are in a better position to reinforce one another's
pro-social behavior, and so may be expected to resist ``invasion'' by
defecting strategies better than when each neighbor interacts with a
different set of others.

The second reason to suspect that network structure should impact
cooperation is that cooperation in networks might be ``contagious.''
Specifically, if A is a conditional cooperator surrounded mostly by
cooperating neighbors, A will cooperate more; but then A's increased
cooperation may cause its remaining neighbors to cooperate more as
well. These neighbors may in turn cause their neighbors to cooperate more
as well, and so on, leading to a cascade of cooperation that sustains
itself over multiple steps. In fact, recently it has been claimed that
cooperation is characterized by a ``three degrees of influence''
rule~\cite{fowler-cascades}, meaning that an individual who increases his
or her level of cooperation can positively impact the contribution of an
individual who is three steps removed from them in the network. Because the
number of individuals who can be reached within three degrees of a
cooperating individual will in general depend on the non-local structure of
the network~\cite{watts-small-world}, the presence of social contagion
would imply that network features other than local clustering should also
impact aggregate cooperation levels.

Although these heuristic arguments suggest that network structure plausibly
impacts cooperation, two other arguments suggest the opposite conclusion.
First, even if it is true that unconditional cooperators will benefit from
preferential interaction and hence network clustering, conditional
cooperation is known to cut both ways, leading as easily to defection as to
cooperation~\cite{axelrod-evolution-book}. In effect, the
assertion that preferential interaction among conditional cooperators will
also aid cooperation makes the additional implicit assumption that
individuals initially cooperate---an assumption that may or may not hold in
practice.  Second, the contagion argument implicitly assumes relatively
``tight'' coupling between neighbors. In coordination games, for example,
paired individuals have very clear incentives to choose actions to
coordinate with their network neighbors.  For example, if A chooses an
action that does not coordinate with a one of its neighbors B, then B will
have a clear incentive to change its action to accommodate A.  If B changes
its action, then another of B's neighbors, say C, who is not directly
connected to A will nevertheless have an equally clear incentive to
coordinate with B as well.  In coordination games, therefore, it is easy to
see how the influence of one player's action can propagate along chains of
intermediaries to affect non-neighbors.  And because conditionally
cooperative strategies have something of the flavor of coordination games,
it is tempting to infer that they lead to the same kind of
contagion---indeed it is precisely this intuition that studies
like~\cite{fowler-cascades} appear to support. However, it is much less
clear that individual strategies for resolving social dilemmas do in fact
exhibit the same kind of coupling as observed in coordination games, or
even should in theory.
%% For example, theoretical work by \cite{bramoulle-network} makes an equilibrium
%% prediction of contributions in a networked public goods setting where the
%% utility function is assumed to be strictly concave, but makes no prediction
%% for linear case that we study here.

In addition to these theoretical arguments, simulation studies of games
over networks have also reached mixed conclusions with respect to the
impact of network structure on contributions.  For example, a number of
simulation studies of social dilemmas on spatial
lattices~\cite{nowak-spatial, may-spatial}, and more recently on
networks~\cite{watts-thesis, eshel-hooligans}, have found that under
certain conditions network structure impacts levels. It should be noted,
however, that all these results depend on numerous modeling assumptions
regarding the behavioral strategies of individual players. Because so many
strategies are conceivable, and because the success of conditional
cooperation depends on what other strategies are present, it is ultimately
inconclusive what can be learned from simulation studies about how real
human players will interact in networks.

Finally, experimental evidence concerning the role of network structure is
also inconclusive.  Although a number of ``networked games'' experiments
have been conducted in recent years using human subjects~\cite{judd-trade,
  kearns-voting, kearns-coloring, cassar-local}, they have generally
focused on other games, like graph coloring~\cite{kearns-coloring},
consensus~\cite{kearns-voting}, economic exchange~\cite{judd-trade}, and
diffusion of social influence~\cite{centola}.  Many of these experiments
have found that network structure dramatically impacts collective behavior,
consistent with the arguments above. Because all these games differ from
one another in subtle but important ways, and because none of them
precisely resemble social dilemmas, it remains unclear how these findings
can be extended to the question of cooperation on networks.  Meanwhile, the
extensive experimental literature that explicitly addresses cooperation has
largely focused on interactions between pairs~\cite{rapaport-conflict}, or
within small, completely connected groups~\cite{marwell-provision,
  ledyard-survey, fehr-punishment}.
%Perhaps the most closely related experiments examine the difference
%between so-called ``partner'' vs. ``stranger'' conditions, where in the
%former condition individuals play with the same partners for multiple
%rounds, whereas in the latter condition they are randomly rematched on
%each round.  Different studies have reached different conclusions
%regarding the difference between contribution levels for the two
%conditions, suggesting that there is no consistent effect
%%\cite{andreoni_croson}.  Once again, however, this result is difficult to
%extend to networks, where individuals play in the equivalent of the
%partner condition, and it is the relationship between partners that is
%different across different network structures.  In spite of considerable
%recent attention paid both to networks, on the one hand, and social
%dilemmas on the other hand, the question of how network structure impacts
%cooperation among humans subjects therefore remains poorly
%understood. Moreover, networked human subjects experiments are difficult
%to conduct, in large part because in order for networks to differ in
%interesting ways they must be sufficiently large; thus each experiment
%requires many subjects to participate simultaneously.
To our knowledge, only one experiment has been conducted to test directly
for the effects of networks structure, by Cassar \cite{cassar-local}, who concluded that
`small-world' networks (i.e. with high local clustering and short global path lengths) support higher
contribution levels in a linear public goods game than randomly connected
networks---consistent with the intuition outlined above. For reasons we
outline below, however, Cassar's findings were ultimately inconclusive.

As a result of the ambiguous and even conflicting conclusions of previous
theoretical, simulation and experimental results, there is a clear need for
clarifying experimental evidence.  The main substantive contribution of
this paper is to investigate the relationship between network structure and
cooperation in a series of networked public goods experiments.  The
experiments we report on were conducted over the World Wide Web using the
popular crowdsourcing platform, Amazon Mechanical Turk
(\url{http://www.mturk.com}).  AMT is a web-based labor market originally
created to facilitate crowdsourcing~\cite{howe} of tasks, called human
intelligence tasks, or HITs, that are easier for humans than for
machines---such as, image labeling, sentiment analysis, or classification
of URL’s.  In addition to its role as a labor market, however, AMT can also
be thought of as a convenient pool of subjects willing to participate in
laboratory-style behavioral experiments.  Mechanical Turk and other
web-based experimental platforms are becoming increasingly popular with
behavioral science researchers, in part because they allow experiments to
be run faster and more cheaply, and in part because they afford access to
potentially a much broader cross-section of the population than is typical
of university-based lab experiments~\cite{mason-watts, ipeirotis-qa,
  paolacci, horton, mason-methods}.  A second contribution of this work, therefore, is to
advance the scope of behavioral experiments conducted on AMT to include
networked games and more generally, games where all players play
simultaneously.

\section*{Results}
We conducted a total of 113 experiments on AMT over a period of 6 months.
In each of these experiments participants played a widely studied
variant of a social dilemma, called a public goods or common pool resource
game~\cite{ostrom-revisiting, ostrom-coping}.  Typically such games last
for a number rounds, where in each round individuals make voluntary
contributions to a common pool. The pool is then augmented in some manner,
reflecting the added benefits of the public good.  After augmentation the
pool is then redistributed to the players, where all players receive an
equal share regardless of their contributions.  Although many specific
variants of this general class of games have been
proposed~\cite{ostrom-coping}, we studied a variant of a standard one in
the experimental literature~\cite{marwell-provision, ledyard-survey,
  mark-group} defined by the payoff function $\pi_i = e_i - c_i +
\frac{a}{n}\sum_{j=1}^nc_j$ where $\pi_i$ is the payoff to individual $i$,
$e_i$ is $i$'s endowment, $c_i$ is $i$'s voluntary contribution, $a$ is the
amount by which collective contributions are multiplied before being
redistributed, and $n$ is the group size. Critically, when $1 < a < n$,
meaning that the marginal per capita return $M = \frac{a}{n}$ lies in the
range $0 < M < 1$ players face a social dilemma in the sense that social
welfare is maximized when all individuals contribute the maximum amount,
but players have a selfish incentive to free ride on the contributions of
others.

\subsection*{Experimental Design}

In contrast with standard public goods games, in which participants'
contributions are shared among members of the same group, here participants
are arranged in a network. To reflect this change, players' payoffs are
subject to the modified payoff function $\pi_i = e_i - c_i +
\frac{a}{k+1}\sum_{j \in \Gamma(i)}c_j$, where in place of the summation
over the entire group of $n$ players, payoffs are instead summed over
$\Gamma(i)$, the network neighborhood of $i$ (which we define to include
$i$ itself), and $k$ is the vertex degree (all nodes in all networks have
the same degree). Therefore, $i$'s contributions are, in effect, divided
equally among the edges of the graph that are incident on $i$, where
payoffs are correspondingly summed over $i$'s edges.  Aside from this
change, our experimental design was kept as similar as possible to previous
work, in order to make comparisons possible. Specifically, we ran each
experiment for 10 rounds, where the first two rounds each lasted 45 seconds
and all subsequent rounds lasted 30 seconds. In each round, each
participant received $e = 10$ after which they were required to nominate a
contribution $0 \leq c_i \leq e$ to the common pool.  The pool was then
augmented and then redistributed to the players, where all players received
an equal share regardless of their contributions, as described in the
payoff function above. At the end of each round, each player received the
following information, which is identical to the information given to the
players in~\cite{fehr-punishment}: (a) their contribution for that round,
(b) the contributions of each of their neighbors for that round, and (c)
their own cumulative payoff up to that point. The information visible to
players is shown in Figure 1.

%Figure 1
\begin{figure}[t]
  \includegraphics[width=\textwidth]{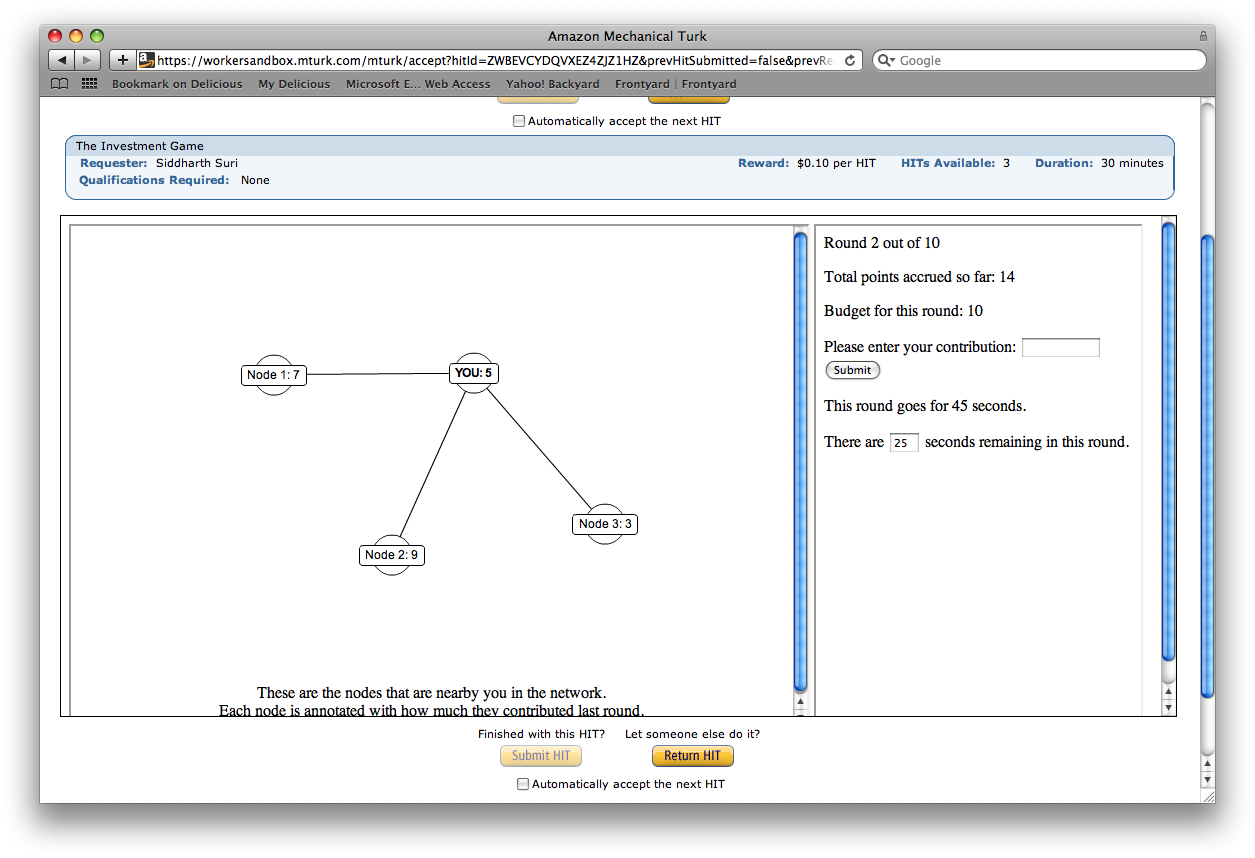}
  \caption{{\bf Screen shot of the experiment.}}
  %\label{Figure_label}
\end{figure}

%Figure 2
\begin{figure}
  \includegraphics[width=\textwidth]{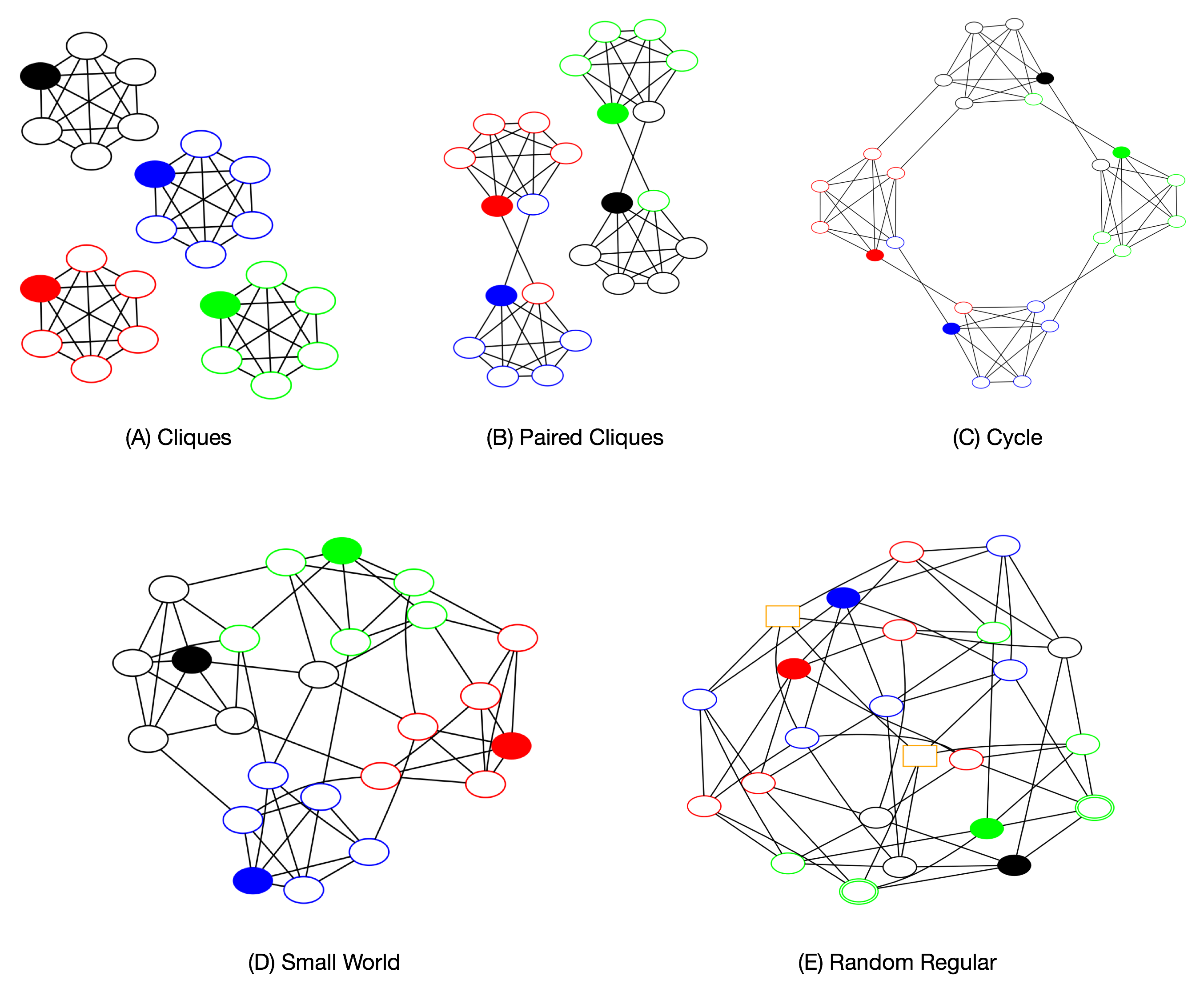}
  \caption{{\bf The five networks used in the experiment.} (A) four cliques of six
    players each; (B) two connected components of twelve players
    constructed by choosing one pair of players in each of two of the
    cliques in A, and swapping partners; (C) cycle of near-cliques
    constructed by choosing a pair in each of the four cliques in A and
    deterministically swapping an edge with a pair from another clique so
    as form a cycle; (D) ``small world'' type network formed by swapping
    four randomly chosen pairs of edges from C; (E) a random regular graph
    in which all nodes have the same degree $k=5$. In all cases, the filled in
    nodes were used as seed nodes in the intervention experiments (see
    text for details). Each seed node is color-coded, and nodes connected
    directly to a given seed are outlined with the same color.  All nodes
    in all networks are directly connected to exactly one seed node, except
    for Random Regular where two nodes are each directly connected to two
    seed nodes (green double circles) and two nodes are not directly
    connected to any seed node (orange rectangles). }
  %\label{fig:networks}
\end{figure}

As shown in Figure 2, we chose networks that spanned a wide range of
possible structures between a collection of four disconnected cliques at
one extreme, and a regular random graph at the other. All networks
comprised $n=24$ players, each with constant vertex degree $k=5$; however,
they varied with respect to three frequently studied structural parameters,
summarized in Table 1: (a) the clustering coefficient $C =
\frac{2}{n}\sum_{i=i}^n \frac{K_i}{k(k-1)}$ where $K_i$ is the number of
completed triangles in node $i$'s neighborhood; (b) the average path length
$L=\langle d_{ij}\rangle$ where the average distance between all pairs of
nodes is taken over each connected component; and (c) the diameter $D=\max d_{ij}$, which
is the distance between the farthest two nodes.  The
clustering coefficient of node $i$ is computed by dividing the number of
triangles incident on $i$ by the number of triangles possible given $i$'s
degree. The clustering coefficient of a network, which is the average clustering coefficient
over all nodes, is therefore a local measure of structure that captures the extent to
which the neighbors of $i$ are also neighbors of each other. The average path length and diameter, by contrast, are global network
measures that quantify the extent to which effects can propagate along
chains of network ties.

\begin{table*}
  \caption{Properties of the five network topologies}
  \begin{tabular}{cccccc}
    & Cliques & Paired Cliques & Cycle & Small World & Random Regular \\ \hline
    Clustering Coefficient (C) & 1.00 & 0.80 & 0.60 & 0.41 & 0.09 \\
    Average Path Length (L) & 1.00 & 1.81 & 2.54 & 2.23 & 2.01 \\
    Diameter (D) & $\infty$ & $\infty$ & 5 & 4 & 3 \\
    Return on Investment (ROI) & 1.04 & 1.09 & 1.38 & 0.80 & 1.00 \\ \hline
  \end{tabular}
  %\label{tbl:stats}
\end{table*}

In spite of these structural differences, we note that from the perspective
of the players, all positions in all networks will seem
indistinguishable---players always see themselves interacting with a local
network of five others, as in Figure 1.  Why then, might we expect the
network structure to make any difference? The answer is that although
players always play with $k$ neighbors, the relationship between their
neighbors changes as a function of the network.  When the network in
question is a clique---a set of $k+1$ nodes in which every node is
connected to every other---our formulation reduces to the standard design
in which the group size $n=k+1$.  For a general network, however, $n$ and
$k$ can be specified more or less independently (except that $n \geq k+1$),
and the connectivity between an individual $i$'s neighbors can also vary
dramatically.  In a clique, that is, every neighbor of $i$ is connected to
every other neighbor, whereas in a random graph, $i$'s neighbors will be
connected to each other with probability roughly $k/n$ which tends to $0$
when $n \gg k$.  We note that our design differs from previous studies that
have compared so-called ``partner'' vs. ``stranger''
conditions~\cite{fehr-punishment}, where in the former condition
individuals play with the same partners for multiple rounds, whereas in the
latter condition they are randomly rematched on each round.  In our design,
individuals always play with the same people as in the partner condition.
It is the relationship between partners that is different across different
network structures.  If the ``reinforcement'' hypothesis, outlined above,
is correct, therefore, the actions of an individual's neighbors ought to be
dependent on the actions of their neighbors, and hence the experience of
the focal individual will depend on the density of interaction between his
or her immediate neighbors.  Likewise, if the ``contagion'' hypothesis is
correct, the focal individual's experience will depend in addition on the
actions of individuals by two or more steps away.  Thus our choice of
topologies was specifically designed to highlight the importance both of
local reinforcement and contagion.

\subsection*{Recruiting and Retention.}  
The Amazon Mechanical Turk (AMT) community comprises two
types of actors: requesters and workers.  Requesters can be individuals or corporations, and
can list jobs along with a specified compensation.  Workers, also known as ``turkers,''
are paid by requesters to complete individual tasks. 
When choosing a task to
work on, workers are presented with a list of jobs that are subdivided into
HITs. Each job contains the title of the job being offered, the reward
being offered per HIT, and the number of HITs available for that
job. Workers can click on a link to view a brief description of the task,
or can request a preview of the HIT.  In our case, we posted the
experiment as a HIT and recruited workers as subjects to do the
experiment. After seeing the preview, workers could
choose to accept the HIT, at which point the work was officially assigned to
them and they could begin completing the task.  HITs range widely in size and
nature, requiring from seconds to hours to complete, and compensation
varies accordingly, but is typically on the order of \$0.01-\$0.10 per HIT.
Currently, several hundred requests may be available on any given day,
representing tens of thousands of HITs (i.e. a single request may comprise
hundreds or even thousands of individual HITs). AMT also provides a
convenient API for transferring payments from requesters to workers.

Although AMT and other web-based experimental platforms are becoming
increasingly popular with behavioral science researchers, the bulk of
previous work has relied on experimental designs that are asynchronous, in
the sense that they do not require a large group of subjects to participate
at the same time. In~\cite{salganik-inequality}, for example, participants
arrived sequentially, and only saw information about the behavior of
previous participants, while in~\cite{ahn-gwap}, at most pairs of
participants were required to be present simultaneously.  In our
experiment, however we required all players to participate
simultaneously---a problem that is solved in physical labs by announcing
official start times and supervising experiments with trained proctors. To
resolve this problem, we instituted a number of web-specific experimental
procedures, as described next and in more detail in \cite{mason-methods}.

\subsubsection*{The Waiting Room} Because it was impossible to assure that participants arrived at precisely the same time, 
and also because different participants required more or less time to read the instructions 
and pass the quiz (see below), we created a virtual ``waiting room,'' similar to~\cite{egas-riedl}. Once they had accepted 
the HIT and passed the quiz, participants saw a screen informing them that the experiment had 
not yet filled, along with how many remaining players were required.  Once all positions had 
been filled, participants in the waiting room were informed that the game was about to commence.  

\subsubsection*{The Panel} In a series of preliminary experiments, we learned that simply posting the HIT on AMT was 
insufficient to fill networks of size $n=24$ in a reasonable time,
resulting in participants abandoning the waiting room and the HIT being
terminated.  To alleviate this problem, we ran a series of experiments with
$n=4$, for which waiting times were reasonable, and then at end of each
experiment, allowed participants to opt-in to being notified of future runs
of our experiment.  In this manner, we created a standing panel of 152
players who had played previously and who understood the instructions
(i.e. they qualified as “experienced” players, consistent with previous
work~\cite{ledyard-survey}). All 113 experiments reported here were
conducted using this panel, the self-reported demographic composition of
which is reported in Table 2. The evening before any experiments were
to be held, we sent messages to the panel (via the AMT API), informing them what
time the experiments would be, typically at 11am,
1pm, 3pm and 5pm EST, although other times of day were used in a few
instances.  We also posted the time of the next days experiments on
\url{turkernation.com}, a bulletin board site for turkers.  At the announced
times, participants would log in to AMT, where the first 24 players to read
the following instructions and pass the quiz at the end of it were allowed
to enter the experiment.

\begin{table}
	\caption{
	\bf{Self reported demographic information of panel members}}
	\begin{tabular}{ccc} \hline
		Gender 		& 	Male 			& 	61.8 \\
	   			 		& 	Female 			& 	35.5 \\
	   		     		& 	Did Not Answer 	& 2.7 \\ \hline
	    Average Age 	&					& 32 \\ \hline
		Highest degree or level of school completed 	& High School 	& 21.1 \\
														& Associates   	& 9.2 \\
														& Bachelors 	& 42.1 \\
														& Masters      	& 18.4 \\
														& Doctorate     & 3.9 \\
														& Professional 	& 3.9 \\
														& Did Not Answer& 1.4 \\ \hline
	   Race			& Asian          						&26.3 \\ 
						& Black or African American 			&1.3 \\
						& American Indian or Alaskan Native 	&0.7 \\
						& White                            	 	&69.7 \\
						& Did Not Answer						&2.0 \\  \hline
	   Marital Status	& Divorced 			&4.6 \\
						& Now Married 		&42.1 \\
						& Never Married 	&49.3 \\
						& Separated 		&2.0 \\
						& Did Not Answer	&2.0 \\ \hline
       Total Annual Household Income 	& $<$ 10k 			& 13.8 \\
										& 10k - 20k 		& 13.2 \\
										& 20k - 30k 		& 9.9 \\
										& 30k - 40k 		& 12.5 \\
										& 40k - 50k 		& 15.1 \\
										& 50k - 60k 		& 7.2 \\
										& 60k - 70k 		& 5.3 \\
										& 70k - 80k 		& 4.6 \\
										& 80k - 90k 		& 2.0 \\
										& 90k - 100k 		& 2.6 \\
										& 100k - 150k 		& 6.6 \\
										& $>$ 150k 			&5.9 \\
										& Did Not Answer	&1.4 \\ \hline
\end{tabular} 
% \begin{flushleft}Table caption
% \end{flushleft}
\label{tab:label}
\end{table}

\subsubsection*{Handling Dropouts} In spite of these precautions, individual participants would occasionally fail to enter a 
contribution on one or more turns, or leave the game entirely. In rare
instances, a participant who had accepted the HIT and passed the quiz did
not participate at all in the game.  To handle these circumstances, we
adopted the following rules: 1) If a player had entered at least one
contribution, and if they subsequently failed to enter a contribution, the
system would automatically enter the same contribution as their previous
round. 2) If a player did not enter an initial contribution, the system
would random choose a contribution of either 0 or 10 for that player (roughly 70\% of the contributions during round 1 where either 0 or 10).  To
avoid biasing our results, we only used data from a given realization if at
least $90\%$ of the contributions in the entire experiment were actually
made by human
players.  %% It turned out that by enforcing this
%% requirement, for any player in any experiment, at least half of the
%% contributions were made by that player, and even then this only occurred in
%% a handful of experiments.  The overwhelming majority of experiments had
%% very few dropouts at all.
% As a result we discarded 5 of the n=4 realizations with repeat players, 17 of the n=4 realizations with fresh 
% players, 9 of the n = 24 human realizations (no seed players), 8 of the n=24 experiments with seed 
% players in the cover arrangement, and 4 of the n=24 experiments with seed players in the concentrated arrangement.

\subsection*{Calibrating the AMT population}

Before proceeding with our main results, we first address two legitimate
sources of skepticism regarding web-based experiments.  First, subjects
playing at home or at work may behave systematically differently from those
playing in a physical lab; thus the results obtained in a web environment
may not be comparable to those obtained in lab-based studies.  To address
this issue, we conducted a series of 24 preliminary experiments that were
designed to replicate the conditions of a previous lab-based
study~\cite{fehr-punishment}. Specifically, we arranged the players in
completely connected groups (cliques) of $n=4$ (equivalent to $k=3$) and
set $M=0.4$.  One difference between our design and~\cite{fehr-punishment}
was that per-round endowments in our experiment were 10 points, instead of
20.  Normalizing for these different endowments, however, Figure 3A shows
striking agreement between the two sets of results, where we note that
qualitatively similar average contribution levels have also been found in
other experimental studies~\cite{ledyard-survey}.  A second issue is that
the compensation rates in AMT are substantially lower than in traditional
lab experiments; thus one might suspect that subjects are correspondingly
less motivated to play seriously.  Previous studies such
as~\cite{camerer-incentives} have shown that for these types of economic
experiments, paying a low or high rate does not have a large impact on
results as long as the payoff amount has a nonzero dependency on
performance.  Nevertheless, we conducted an additional series of 16
experiments which alternated the compensation between \$0.01 per point and
\$0.005 per point (participants were also paid a fixed up-front fee of
\$0.50 for accepting the task and passing the quiz).  As Figure 3B shows,
contribution levels for both compensation levels were similar, which is
also consistent with prior work~\cite{camerer-incentives}. We therefore
conclude that neither compensation rates nor context significantly affected
the behavior of subjects in our games, relative to previous studies.  Thus
reassured, we now proceed to discuss our main results, which concern
behavior on networks.

%Figure 3
\begin{figure}
  \centering
  \includegraphics[width=\textwidth]{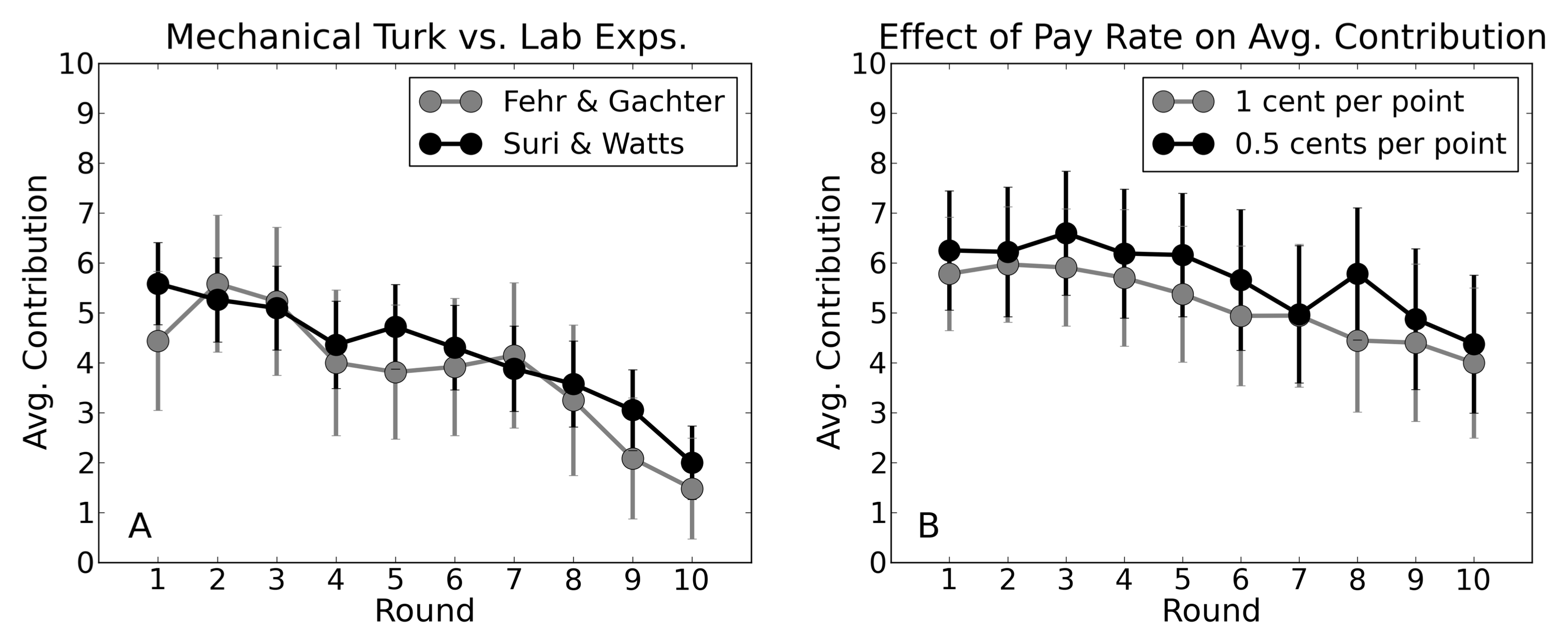}
  \caption{{\bf Calibrating the AMT platform.} (A) Comparison of
    contributions for identical linear public goods games conducted on
    Amazon Mechanical Turk and in a physical lab~\cite{fehr-punishment}.
    (B) Contributions for different compensation levels.  In both panels
    error bars indicate 95\% confidence intervals.}
  %\label{fig:calibrate}
\end{figure}

% Results and Discussion can be combined.
\subsection*{Testing for Effects of Network Structure}

%Figure 4
\begin{figure}
  \includegraphics[width=\textwidth]{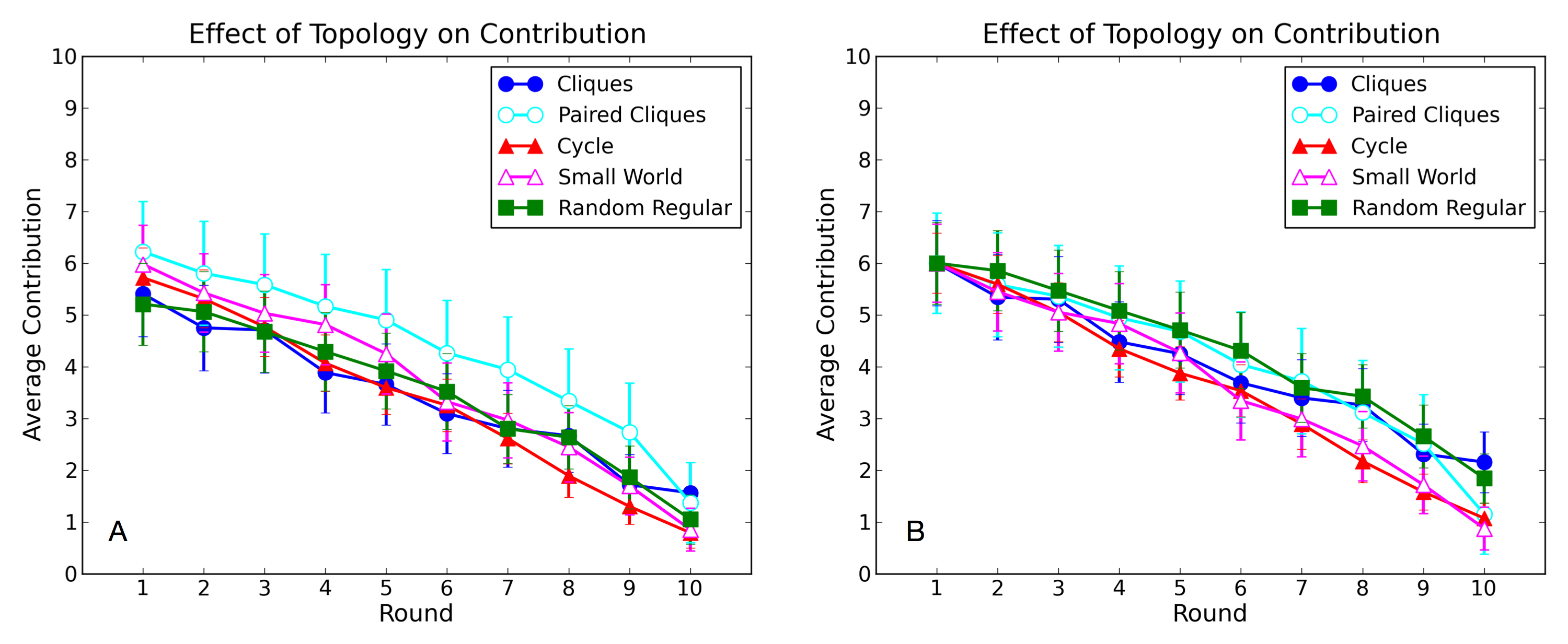}
  \caption{{\bf Average contributions per round for each of the five network
      topologies shown in Figure 2.} (A) Raw contributions. Error bars indicate 95\%
    confidence intervals.(B) Contribution curves shifted vertically so that they all start at the same point.}
  	%\label{fig:topology}
\end{figure}

In the first set of network experiments all positions in the network were
filled by human players recruited from AMT.  Because individual
contributions tended to vary considerably from one experiment to the next,
and different players were likely to play at different times of day, we
conducted multiple realizations of the experiment for each topology (see
Table 3). The order and timing of experiments was randomly varied between
realizations.  In total, we conducted 23 experiments over a period of 8
weeks.  Figure 4A shows the average contribution for each round, for each
of the five topologies.  Visually, the average contribution follows a very
similar pattern regardless of the network topology.  This result is
confirmed by a Kruskal-Wallis test~\cite{siegel-stats} on the five
distributions (one for each topology) of contributions for each round,
which found no significant differences (the smallest P-value is for round
8: H=6.43, df=4, P=0.17). Figure 4A also shows that contribution curves
that start higher, relative to other curves, tend to stay above the other
curves over the course of the experiments; yet, clearly the first round
contributions are random and unrelated to the topology of the network.  To
see the differences between topologies more directly, therefore, Figure 4B
shows the same contribution curves as in 4A, but shifted vertically in
order that they have the same initial value. As can be seen, eliminating
these initial difference further diminishes the already small differences
between topologies.

\begin{table*}
  \centering
  \caption{The breakdown of realizations per topology is given. The larger
    number of cycle topology experiments was due to the 
    presence of two outliers: experiments in which uncharacteristically
    high contributions were registered. The effect of these outliers was to
  greatly increase the size of the error bars for that topology, thus more
  realizations were required}
%%   \begin{tabular}{c|ccccc}
%%     & Cliques & Paired Cliques & Cycle & Small World
%%     & Random Regular \\ \hline 
%%     All Human & 4 & 3 & 8 & 4 & 4 \\
%%     Cooperative Seeds & 3 & 2 & 4 & 2 & 2 \\
%%     Defecting Seeds & 2 & 2 & 9 & 2 & 2 \\
%%     Concentrated Seeds & N/A & 4 & 5 & 5 & 6\\
%%     \end{tabular}
  \begin{tabular}{cccccc}
    & & Paired  &  & Small 
    & Random  \\ 
    & Cliques & Cliques & Cycle &  World
    &  Regular \\ \hline 
    All Human & 4 & 3 & 8 & 4 & 4 \\
    Cooperative Seeds, Cover & 3 & 2 & 4 & 2 & 2 \\
    Defecting Seeds, Cover & 2 & 2 & 9 & 2 & 2 \\
    Cooperative Seeds, Concentrated  & N/A & 4 & 5 & 5 & 6 \\ \hline
    \end{tabular}

\end{table*}

%Figure 5
\begin{figure}
  \centering
  \includegraphics[height=\textheight]{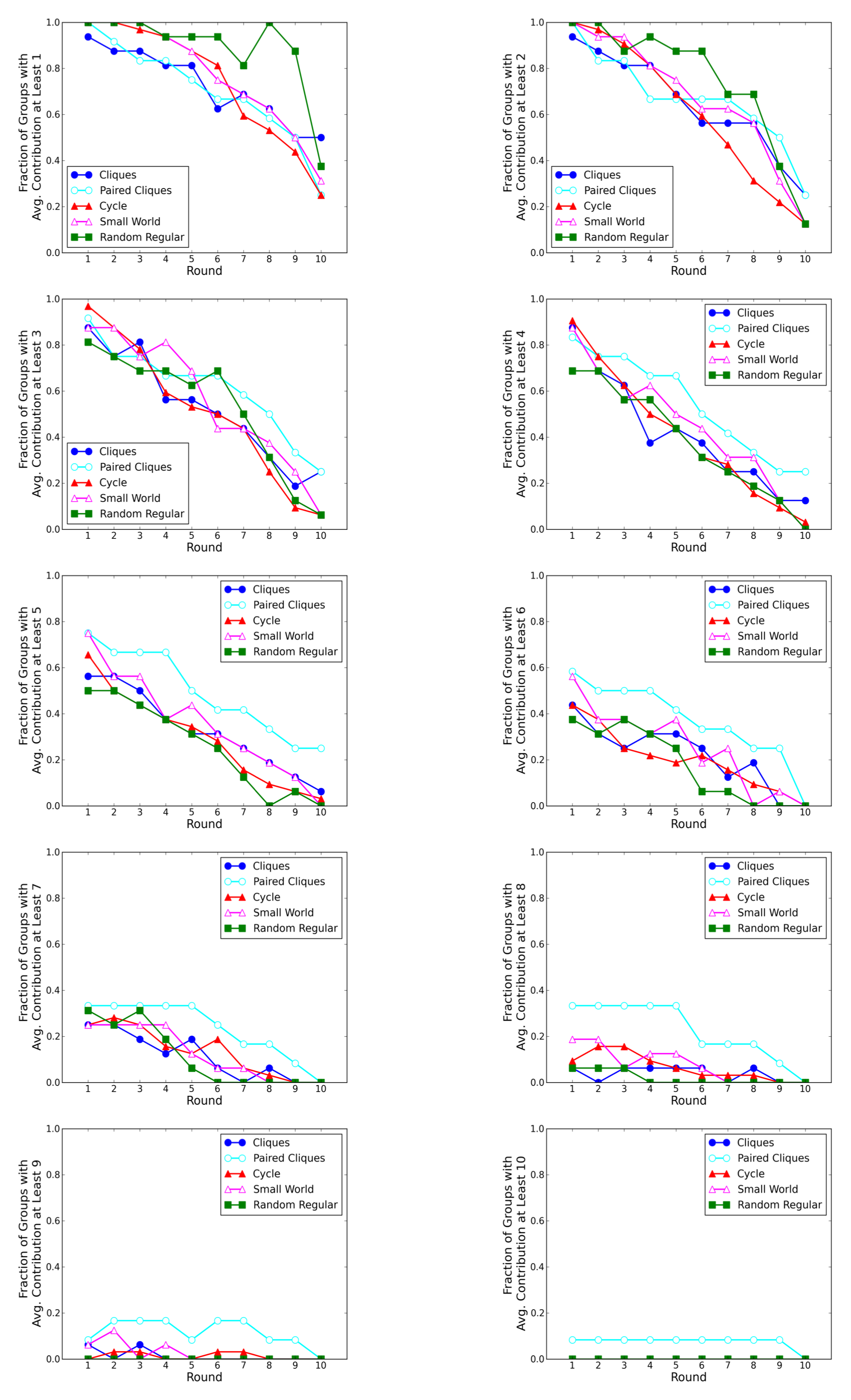}
	\caption{{\bf Fraction of groups with average contribution at least $X$, where $1 \leq X \leq 10$.}}
	%\label{Figure_label}
\end{figure}

%Figure 6
\begin{figure}
  \centering
  \includegraphics[height=\textheight]{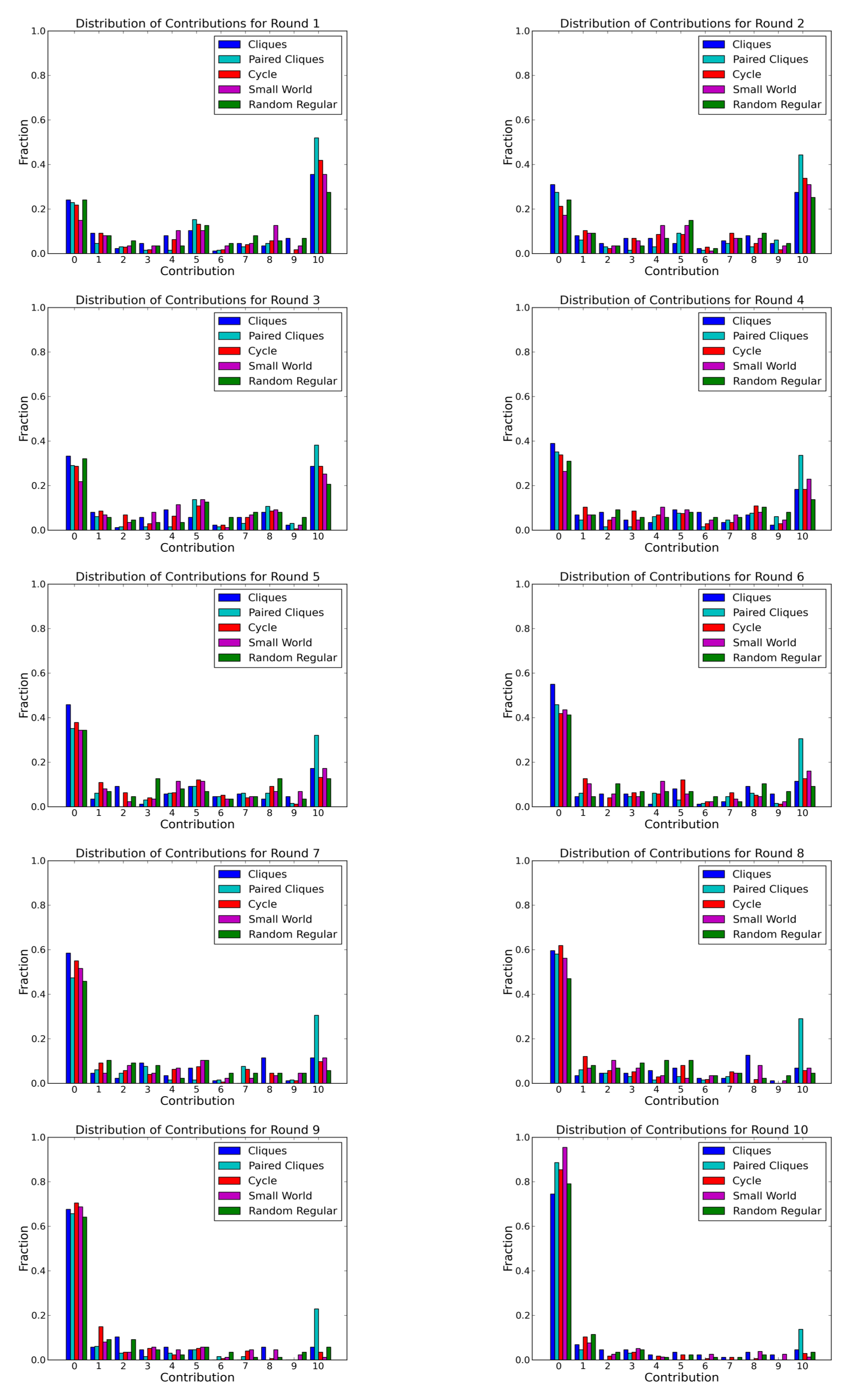}
	\caption{{\bf Distributions of individual level contributions across topologies.} The distributions vary only slightly as the topology is 
	changed.  One realization of the Paired Cliques topology was an outlier; it had a higher then normal number of full contributors.}
	%\label{Figure_label}
\end{figure}

In addition to considering differences in aggregate contributions,
we also checked for differences between topologies both at the level of
individual nodes, and for individual ``groups'' defined as the nodes that are
assigned to the same cliques in topology 1 (see colors in Figure 2). 
As Figure 2 indicates, these groups become progressively less meaningful as 
the clustering coefficient diminishes: in the Random Regular topology, two nodes 
in the same group (same color) are no more likely to be connected than nodes of 
different groups. In spite of these topological differences, however, Figure 5 
indicates that they do not impact contributions; specifically, the fraction of groups 
contributing at least X in a given round is similar for all topologies, and over all rounds. 
Finally, Figure 6 shows the full distribution of individual level contributions for the 
five topologies (color coded) over all ten rounds. Although all distributions change 
dramatically over the course of the game, reflecting the average decline in contributions 
seen in Figure 4, the changes are similar for all topologies.  Thus we conclude 
that topology does not exert a noticeable impact on contributions at any level: individual, group, or aggregate.

\subsubsection*{Comparison of Results with Cassar (2007)}

Cassar~\cite{cassar-local} conducted a total of 11, 18-player prisoner
dilemma experiments on networks of players, where the networks were varied
between the following three topologies: a ``local'' network on which
individuals were arranged on a cycle, and each individual was connected to
their two nearest and two next-nearest neighbors (i.e. $k=\bar{k}=4$ for
all nodes); a ``small-world'' network in which a small fraction of the
edges in the cycle were rewired (hence $\bar{k}=4$ , but individual $k$
varied); and a ``random'' network in which individuals were randomly
connected (again, $\bar{k}=4$ , but individual $k$ varied).  Three
realizations of each topology were tested; thus clustering coefficient
varied between $0.06 \leq C \leq 0.5$, depending on topology, and path
length varied between $2.03 \leq L \leq 2.67$, where the local topology had
the highest $C$ and $L$, the random topology had the lowest, and the
small-world topology was intermediate. Cassar found that cooperation in the
small-world topology was significantly lower than either the local or the
random topology (Table 5, p. 224 in~\cite{cassar-local}). She also found
that in a logit model, the terms for C and L were negative and positive
respectively, and both were significant (Table 10,
p. 227~\cite{cassar-local}).

At first glance, these findings appear to contradict our own; however, we
note that the differences reported as significant in Cassar’s Table 5 are
between cumulative contributions, over the 80 rounds of the
experiments. Yet as noted above, and also by Cassar (see her
Footnote 13), if the contributions in one realization start at a higher level
than other realizations, they tend to stay above the
other realizations for the duration of the experiment.  This suggests
that contributions across consecutive rounds are unlikely to be
independent. Combining contributions over many rounds therefore
artificially amplifies the differences, leading to the appearance of
statistical significance where none may exist.  In fact, as Table
5 in~\cite{cassar-local} itself makes clear, the final (and also average)
difference between topologies is roughly the same as the initial difference
(period 1-20); thus essentially all of the difference can be explained in
term of initial contributions, which are by construction unrelated to the
network topology.  Second, the significance of the NetworkClustering and
NetworkLength coefficients in the PD1 logit model (Table
10 in~\cite{cassar-local}) is marginal and disappeared when other factors, such
as the $\%$ cooperation in the previous experiment (PD2) or dummy variables
for the session (PD3) were included.  If simply controlling for the session
in which a game was conducted eliminates the significance of a coefficient,
then it would seem that any claims to significance ought to be regarded
with caution. On closer inspection, therefore, Cassar's results are probably 
consistent with ours---that, is differences in contribution levels between network structures are not significant.  
% Noting that our experiments 
% involve more realizations (23 vs. 11) of larger networks (24 vs. 18) over a wider range of C and L, we 
% therefore are led to conclude that neither clustering nor path length appears to have a significant impact on cooperation.

Although Cassar's results on how network structure impacts contribution
levels in public goods games may be ambiguous, they do support our claim
that the theoretical arguments above~\cite{watts-thesis, nowak-spatial,
  may-spatial, eshel-hooligans} have led researchers to suspect that
network structure should matter.  Specifically, intuition and simulation
results suggest that when conditional cooperators are allowed to interact
preferentially, i.e.\ in networks that exhibit high clustering, they ought
to reinforce each other, thereby sustaining higher contributions for longer
than in randomly connected networks which have low clustering.  Likewise,
the contagion argument suggests that clusters of high contributors ought to
exert a positive impact on the contributions of neighbors who are not in
the cluster, thereby promoting the spread of cooperation.  If in fact,
network structure does not impact contributions, then one or both of these
two arguments must be invalid. To differentiate between these possible
explanations, we conducted two additional series of experiments, which we
describe in turn.

\subsection*{Testing for Conditional Cooperation.}
%%DW to SS: I modified the sentence preceding the discussion of deception
In the first series, comprising 30 experiments over 4 weeks, we followed
the same design as above, but with the key difference that in each
experiment four nodes were selected, one from each group (indicated with a
filled circle in Figure 2), and their contributions were all artificially
fixed either at 10 (the ``cooperative'' condition) or 0 (the ``defection''
condition) for all rounds. We emphasize, that these players were played by a computer,
not by subsidizing real players, where we did not explicitly disclose to
subjects that their neighbors might not be played by other human
players. Behavioral scientists of different traditions have varying
attitudes with respect deceptive manipulations: experimental economists
view them as unacceptable in principle, whereas psychologists practice them
when the research benefit outweighs any harm caused to subjects.  In our
case, subjects were exposed to minimal harm; thus we viewed the benefit of
being able to establish clear causal relations as justifying the
manipulation. 
Potentially the issue could have been
avoided by including a statement in the instructions to participants to the
effect that ``from time to time certain positions may be played by
automated agents rather than humans.'' However, we do not believe that the inclusion
of such a statement would have affected the results.

Following the above procedure, we were able test the conditional cooperator
hypothesis by directly measuring the positive/negative influence of
unconditional cooperators/defectors on their immediate neighbors. We note
that with the exception of the random regular network, the seed players
were arranged in order to cover the network, meaning that each human player
was adjacent to precisely one seed player; in addition, each human player
was connected via two-step paths to all four seed players (in the random
regular case, a perfect cover arrangement did not exist for the selected
network; thus a close approximation was used instead). An advantage of this
arrangement, which we call the ``cover'' condition, is that all human
players were subjected to the same experimentally manipulated influence,
both direct and indirect.

%Figure 7
\begin{figure}
  \includegraphics[width=\textwidth]{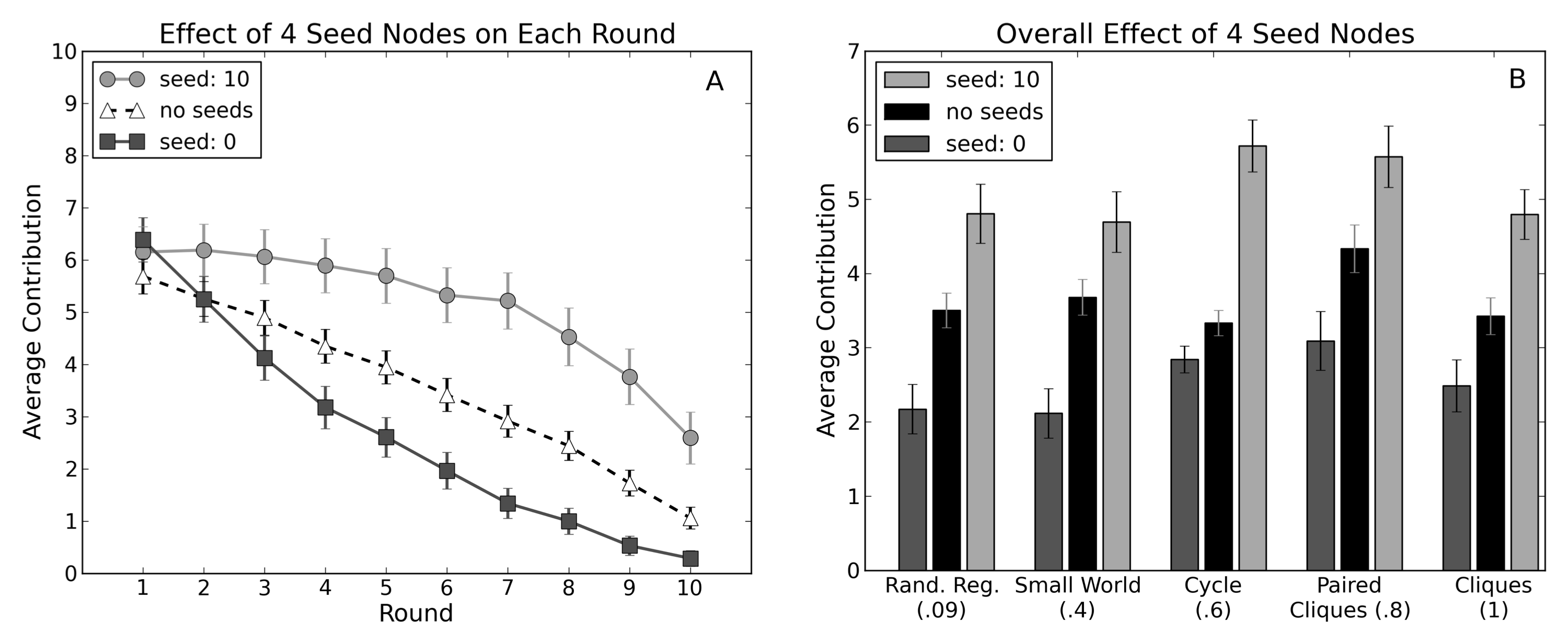}
  \caption{{\bf Contributions for cover-seed experiment.} (A) Average contribution per round for the cooperating and
    defecting conditions averaged over all realizations and all topologies.
    (B) Overall
    average contribution for each topology under the cooperating, defecting
    and all human conditions. The clustering coefficient for each network
    is listed in parenthesis.  In both panels error bars indicate 95\%
    confidence intervals.}
  %\label{fig:topology}
  %\label{fig:intervention}
\end{figure}

Figure 7A shows that in all topologies, the presence of cooperating seeds
stimulated consistently higher aggregate contributions from the remaining
20 players, while the presence of defecting seeds had the opposite
effect. Possessing a high (or low) contributing neighbor therefore did
increase (or decrease) the average contribution levels; thus our subjects
were indeed behaving as conditional cooperators.  Nevertheless, Figure 7B
shows that the effect of the seed players was not consistently bigger in the
graphs with the highest clustering.  For example the effect of the seed
nodes in the Cliques network, which had the maximum number of triangles
incident on each node, was very similar to the effect of the seeds nodes in
the Random Regular network, which had fewer than 1/10th as many triangles.
This result implies that two nodes that form a triangle with a cooperating (or defecting) seed do
not have an appreciably larger (or smaller) average contribution level then
two disconnected nodes with a cooperating (or defecting) seed neighbor in
common.  Mutual reinforcement of the contributions among the neighbors of a
seed node is largely absent, whether or not there is an edge between the
neighbors.

% Figure 8 (used to be table 4)
\begin{figure}
  \centering
  \includegraphics[width=0.5\textwidth]{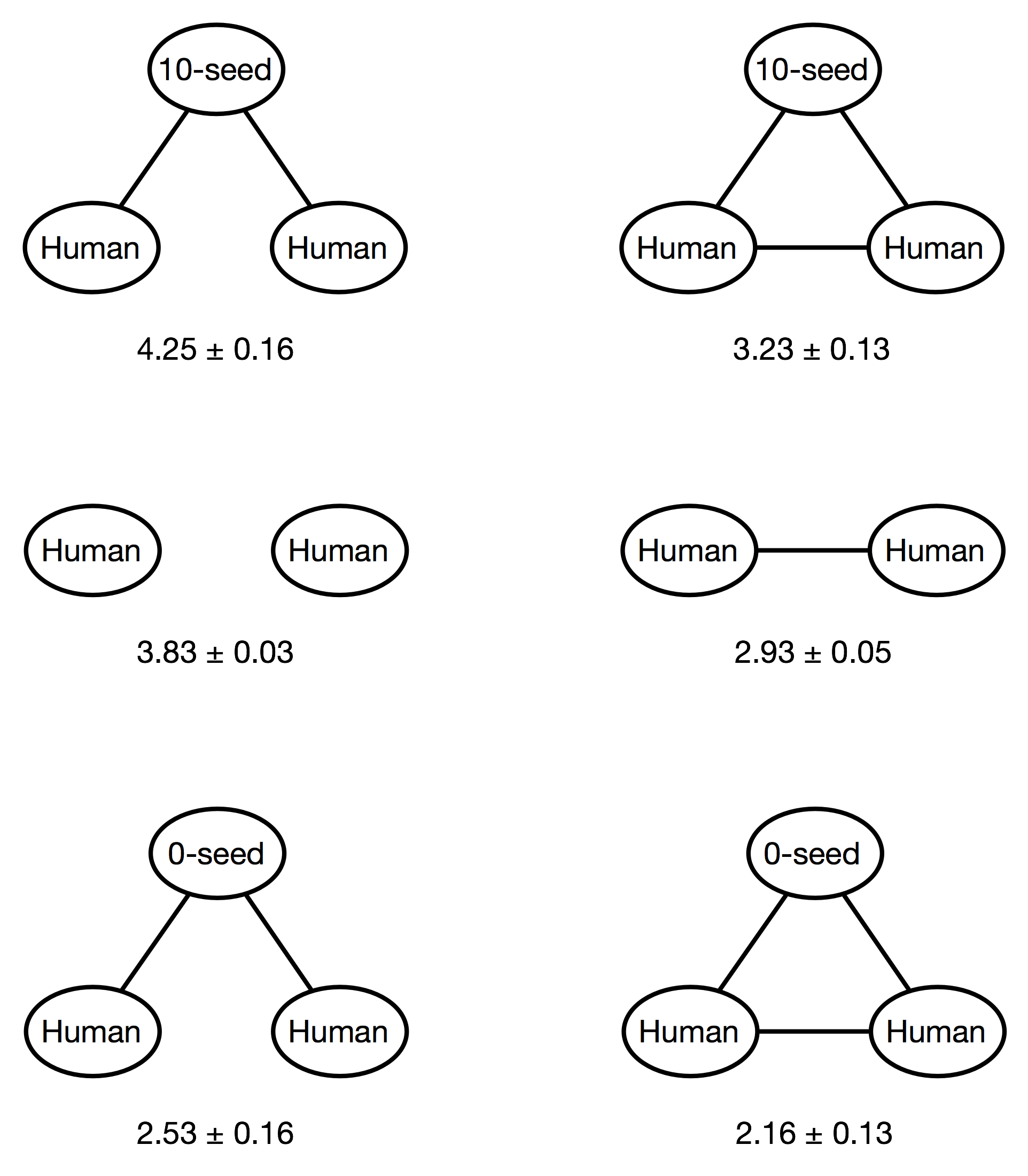}
  \caption{The average pair-wise difference of the human contributions in
    each of the subgraphs pictured.}
  %\label{tbl:coordination}
\end{figure}

% Table 4 became figure 8
Is there in fact any effect of increasing the number of triangles in the
network?  To answer this question, Figure 8 compares the difference in
contributions of pairs of players that (a) are adjacent versus not
adjacent, and (b) share a positive or negative seed as a neighbor versus no
neighboring seed.  Comparing the left column to the right column shows that
adding an edge to a disconnected pair of edges increased the similarity
between their contribution levels.  It also shows that completing a
triangle between two human players and a seed node also increased the
similarity of the contributions of the humans.  Thus, increasing the number
of triangles in the network did indeed increase coordination within the
neighborhoods of the seeds.  We emphasize, however, that the coordinating
influence of triangles cuts both ways by increasing contributions in the
presence of cooperating neighbors and diminishing them in the presence of
defecting neighbors; thus increased coordination among triangles of players 
does not correspond to increased contribution levels. 
Put another way, players do cooperate conditionally, but the negative effects 
of conditional cooperation counteract the positive effects such that the net result is independent of local clustering.

%%DW to SS: edited the end of the previous para

\subsection*{Testing for Contagion.} 
As noted above, another possible explanation
for the lack of impact of network topology on aggregate contributions is
the absence of contagion. That is, even if players do behave as conditional
cooperators, both with respect to the artificial seeds  and also the other
neighbors of seeds, possibly these effects are not strong enough to
propagate beyond the immediate neighborhood of a cooperation seed.
Unfortunately, the above experiment allows us to draw only limited
conclusions regarding contagion.  Since the cover arrangement of seeds
meant that all human players were subjected to the same potential
influence, both direct and indirect, we did not experimentally manipulate
the level of positive/negative influence at different distances from human
players.

%Figure 9
\begin{figure}
  \includegraphics[width=\textwidth]{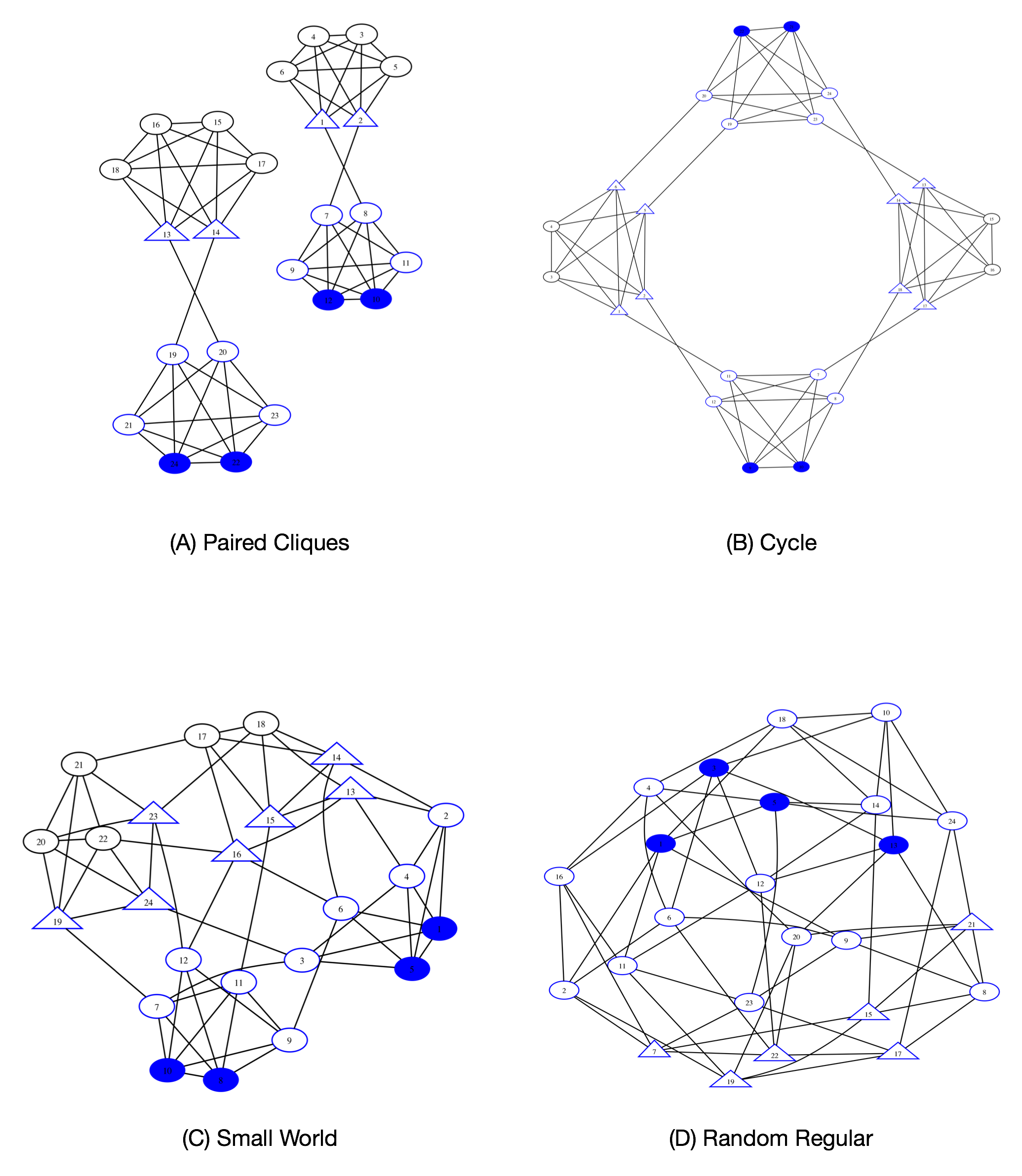}
	\caption{{\bf Cooperating seeds in the concentrated seed
            experiments.}  The blue, filled-in nodes were used as seed
          nodes in the concentrated arrangement (see text for details).
          Oval shaped nodes that are outlined in blue are directly
          connected to at least one seed node.  Triangular nodes are two
          hops from at least one seed node.  In each topology two of the
          seeds in the concentrated arrangement were also seeds in the
          cover arrangement.}
	%\label{Figure_label}
\end{figure}

To further test for the possibility of contagion, therefore, we conducted a
third series of 20 experiments over 2 weeks, in which we kept the number of
unconditionally cooperating seeds constant at four per network (we did not
introduce unconditional defectors in these experiments), but concentrated
them together into two adjacent pairs (see Figure 9).  This arrangement of seeds, which we call the ``concentrated''
condition, therefore exposed some human players to two unconditional
cooperators as immediate neighbors, while others were not exposed to any
seeds directly, but were connected indirectly to the seeds via a human
intermediary.  Since the Cliques topology did not allow for this type of
arrangement  we excluded it from these experiments.  If positive
contagion were present in the network, we would expect to see nodes at
distance two from the seeds increase their contributions relative to the
all-human (i.e.\ no seeds) condition. Moreover, the premise of conditional
cooperation would also lead us to expect that immediate neighbors would
increase their contributions relative to the cover-seed condition.

%Figure 10
\begin{figure}
  \includegraphics[width=\textwidth]{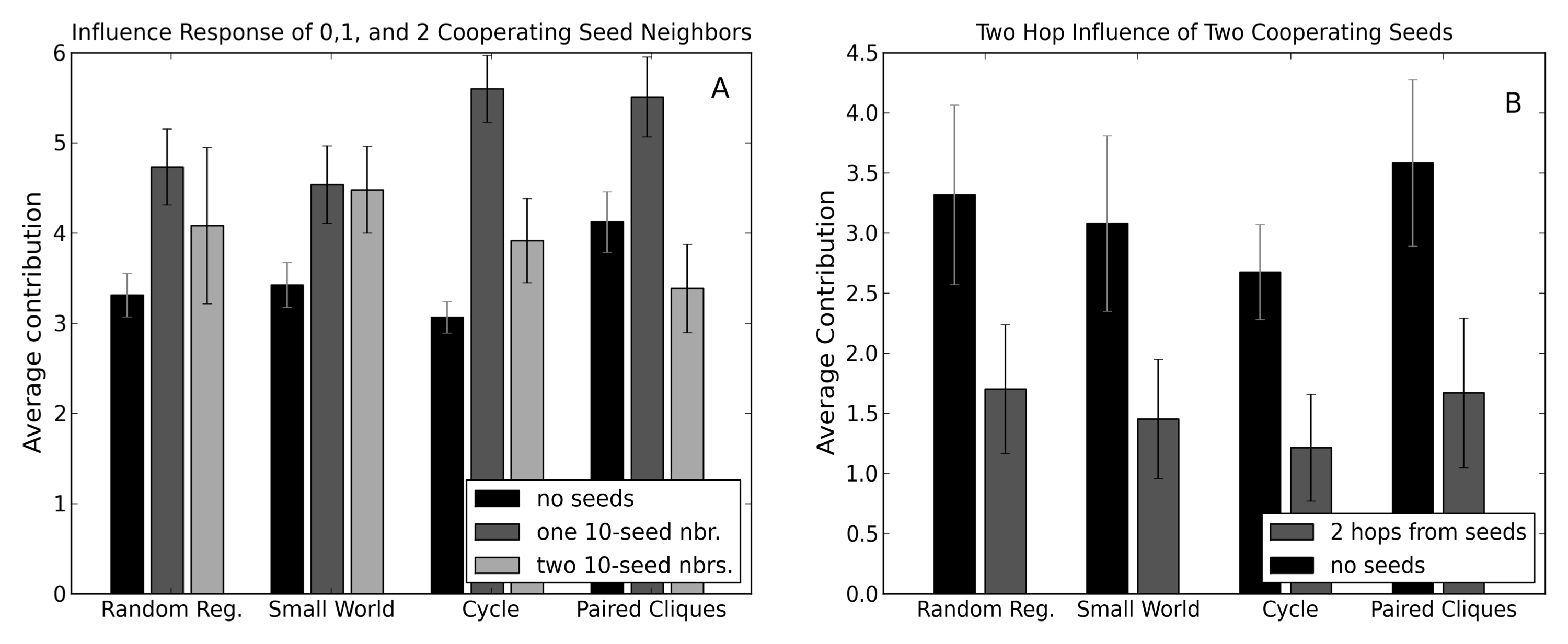}
  \caption{{\bf Contributions for concentrated-seed experiment.}(A) The average contribution of human players neighboring 0, 1
    or 2 cooperating seed nodes.  (B) The average contribution of the human
    players 2 hops from 2 seed nodes compared to the average contribution
    of the corresponding nodes in the all human experiments.  In both
    panels error bars indicate 95\% confidence intervals.}
  %\label{fig:concentrated}
\end{figure}

Surprisingly, our results contradicted both these expectations.  First we
found that nodes who were directly connected to two cooperating seed nodes
contributed more than players who were not
attached to any seed nodes, but less than players who were attached
to only one seed node (both computed from previous experiments) as shown in Figure 10A.  
These results suggest that although many players do respond positively to
the introduction of unconditional cooperators, the presence of too many unconditional cooperators invites free
riding. Conditional cooperation, that
is, appears to be subject to at least two distinct conditions that are in
tension with one another: on the one hand, individuals do not want to
contribute unless others are contributing; but on the other hand, if others
contribute too much, the temptation to free ride overrides their
inclination to reciprocate. In spite of this result, it is nevertheless the case that immediate
neighbors of cooperating seeds did on average contribute more than in the
no-seed condition.  Assuming that the remaining players (i.e.\ at distance
two from the seeds) also cooperate conditionally, one would expect that the
increased contributions associated with the neighbors of a fully
contributing seed would generate contagious effects leading to increased
contributions among these nodes as well. Yet these effects were not
apparent.  Quite to the contrary, in fact, Figure 10B shows that the
two-step neighbors of the cooperating seeds contributed slightly less than
the nodes in the corresponding network positions contributed in the
all-human experiments.

\subsubsection*{Testing for Learning Effects}

%Figure 11
\begin{figure}
  \includegraphics[width=\textwidth]{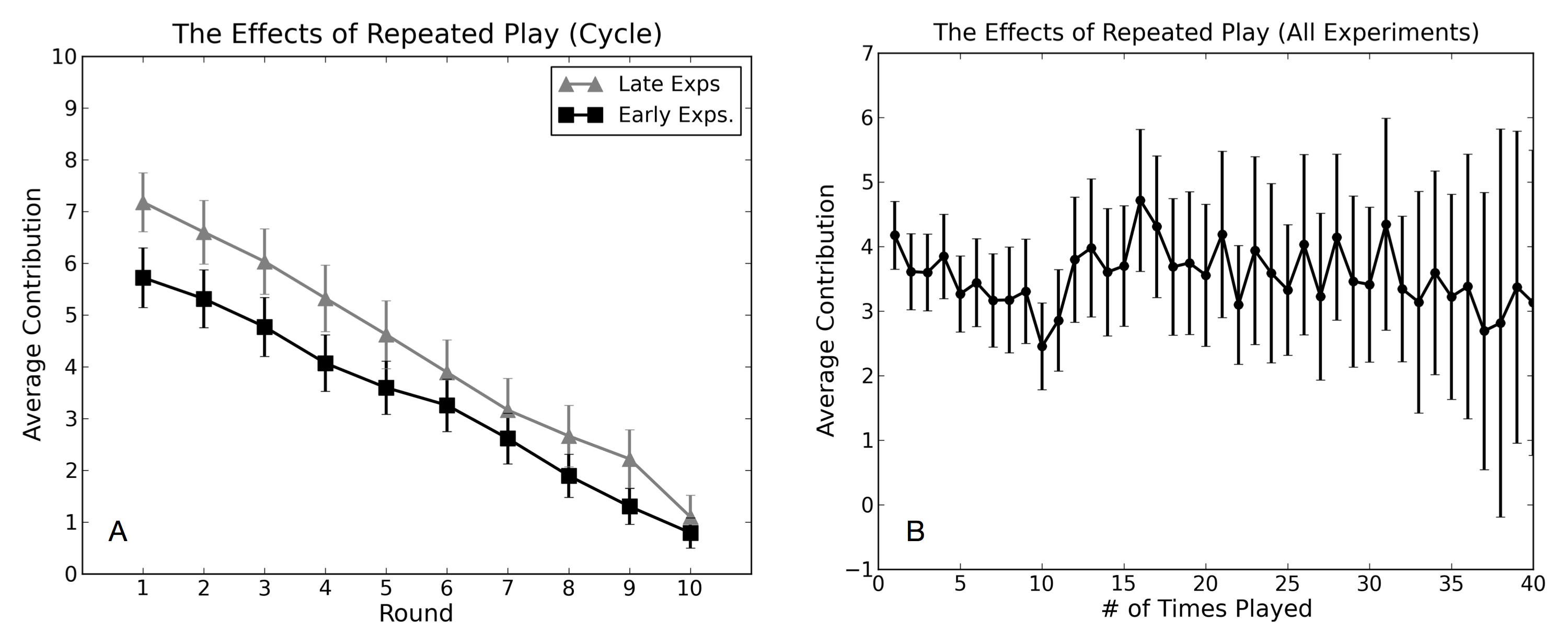}
	\caption{{\bf Checking for the effects of learning.} (A) Comparison
          on the Cycle topology with all human players between experiments
          conducted early in our study and at the end of our study. (B)
          Average contribution levels as a function of how many times an
          individual has played.}  
	%\label{Figure_label}
\end{figure}

To check that this unexpected reduction in contributions did not reflect a
systematic overall shift from higher to lower contributions over the course
of dozens of experiments involving our panel, we ran an additional series
of all-human experiments, finding that average contributions had, if
anything, increased slightly relative to the earlier round of all-human
experiments (see Figure 11A).  We also studied average contributions as a
function of the number of games played by individual subjects, finding that
experienced players who have played as many as 40 games did not contribute
on average, more or less than inexperienced players (see Figure 11B).
Moreover, we tested for selection effects by comparing the complete
history of average contribution levels of those who chose to play many
times to the overall subject pool and did not find a significant
difference.  
Thus we conclude that the reduced contributions observed in
the concentrated seed experiments are not explainable either in terms of a
systemic over-time shift in player behavior, the presence of experienced
players contributing less, or a higher return rate of more cooperative players.
We also note that although experienced players have been used in previous
experiments~\cite{ledyard-survey}, it is unusual to allow subjects to play
upwards of 30 times over a period of months. Previously it has been unclear
whether or not such players would learn over time to play differently,
thereby systematically biasing the results. Figure 11 is therefore
reassuring in that it shows no evidence of such a systematic bias.

\subsubsection*{Comparison of Results with Fowler and Christakis (2010)}

%%DW to SS: Parenthetical note added after Fehr and Gachter cite
Finally, we note that our finding that positive contagion does not occur in
public goods games on networks appears to contradict a recent claim
by Fowler and Christakis~\cite{fowler-cascades} mentioned earlier.  The
authors claim that cooperative cascades
take place on networks of individuals playing a linear public goods game,
and that evidence of contagion persists for up to three steps, leading them
to hypothesize a ``three degrees of influence'' rule.  We note, however, an 
important difference between the networks studied by Fowler and Christakis and those that we have studied here.
Specifically, Fowler and Christakis reanalyzed
data from Fehr and Gachter~\cite{fehr-punishment} (the same results that we replicated in our preliminary experiments described above) in which groups of $n=4$
players were randomly reassigned to new groups after each round.  Whereas in our networks, all individuals appear just once and play with same set of neighbors each turn, in~\cite{fowler-cascades} each individual appears $r$ times (where $r$ is the number of rounds of the experiment) and plays with a different set of neighbors each time. As a result, the measure of network distance in~\cite{fowler-cascades} does not map precisely to the conventional meaning of network distance, which is the meaning that we have adopted here, but rather refers at least in part to the relation between an individual's present and past states.
Although this unconventional definition of distance makes the two sets of results difficult to compare, the main finding in~\cite{fowler-cascades}, that individuals who belonged to higher contributing groups in round
$t-1$ contributed, on average, more in round $t$, seems consistent with our 
observation that initially high contributions tend to persist over time. 
We also note, however, that it was precisely to separate the effects of persistence from ``true'' contagion, in the sense that an effect due to a single individual propagates to a remote individual along a series of network ties, that we designed the concentrated seed experiment. And as the results from that experiment make clear, neither persistence nor even conditional cooperation (as demonstrated in the cover seed design) are sufficient to generate contagion in this sense.  
Given these results, we conclude that although the effects of higher neighbor contributions may well persist for up to three rounds, the most intuitive interpretation of the ``three degrees of influence'' rule---namely that higher contributions spread from individual to individual in a static network for up to three steps---is not supported.
% 
% In the setting where each
% node plays with the same people round after round, we found that there was
% influence on network neighbors---i.e.\ one hop of influence.  Whereas our
% concentrated seeds design was specifically designed to detect contagion
% along network ties, Fowler and Christakis' design was designed to detect
% contagion along a temporal axis.  Thus their results are not necessarily
% inconsistent with ours, even though they reach a very different conclusion.

\section*{Discussion}
Returning to our original motivation, theoretical arguments in favor of an
association between network structure and cooperation invoke two related
ideas: first, that individuals are conditional cooperators, increasing
their contributions in response to the increased contributions of their
neighbors; and second, that positive effects of conditional cooperation
should propagate through the network via a process of contagion.
In this paper, we have tested the effects of network topology on
contribution levels in a standard public goods game, finding no significant
effects.  In addition, we conducted two separate rounds of experiments---one
to test for the presence of conditional cooperation, and the other to test
for the possibility of positive contagion. Although we do find strong
evidence of conditional cooperation, we do not find evidence of positive
contagion in the standard sense of multi-step propagation along a sequence of ties in a static network.

Our explanation for these results is that the theoretical arguments cited
above emphasize the positive aspect of conditional cooperation, yet
conditional cooperation implies not only that players increase their
contributions in response to cooperative neighbors, but also that they
decrease their contributions in response to defecting neighbors. 
Although it is the case that highly clustered networks offer more
opportunities for positive effects to reinforce each other than random
networks, they also offer more opportunities for negative effects to
reinforce each other as well. 
By contrast, in random graphs where there is very little clustering,
neither cooperation nor defection get reinforced and seeds act as influence
blockers preventing either positive or negative influence from propagating
among neighbors.

%% SS to DW:  I just added this paragraph.
As stated in the introduction, in the case of coordination games, if node A
chooses an action that results in a lack of coordination with neighbor B,
then B has a clear incentive to change its action.  In turn, if this
results in a lack of coordination with C which is a neighbor of B and not
A, this can result in contagion.  In a cooperation setting, B need not
change its action in response to A because the incentives do not enforce
such a tight coupling of neighbors actions.  This leads to an
interesting open question---under what theoretical conditions should one
expect to see contagion over networks with fixed neighbors?  In
demonstrating that not all dynamic games on networks exhibit contagion
we hope that our findings will provoke new theoretical hypotheses
along these lines, as well as new experiments to test them.

Moreover, even in the absence of contagion, our observations also show how an outside entity
might stimulate cooperation in a network by subsidizing targeted
individuals to cooperate or by inserting unconditionally
cooperative players into the network.
We emphasize that unlike other known strategies for stimulating
cooperation, such as allowing punishment~\cite{fehr-altruistic} or
reward~\cite{milinski-reputation}, or introducing sanctioning
institutions~\cite{gurerk-advantage}, this mechanism does not change the
game by giving players another action, but instead exploits the network on
which the game is being played.
As Table 2 shows, in the cover experiments the positive intervention was
cost-effective in four out the five topologies.  More specifically, the
expected cost of subsidizing players, i.e.\ the additional contributions of
the four seeds over their average contribution in the no-intervention case,
was less than the total marginal increase in contributions from the
remaining twenty individuals.
These results therefore provide empirical support for
earlier theoretical work~\cite{heal-ids} which proposed that seeding or
inducing cooperation among focal actors may generate positive effects on
the network. 
Our work also suggests where to place the seed nodes for maximum effect.
The absence of positive contagion---along with the
negative marginal effect on neighbors of multiple unconditionally
cooperating seeds---implies that the impact of cooperative seeds is
maximized by spreading them widely across many groups, thereby maximizing
the total number of human players exposed directly to seeds.  

In concluding, we note that in addition to their substantive relevance, the
experiments discussed here also demonstrate the possibility of web-based
behavioral experiments involving the simultaneous presence of many players (see also~\cite{egas-riedl}).  
Although experiments of this nature and scale have been
conducted in physical labs~\cite{judd-trade, kearns-voting,
  kearns-coloring, cassar-local}, web-based ``virtual labs'' exhibit two
important advantages over their physical counterparts: first, experiments
can be run faster and more efficiently (e.g. we ran 113  experiments costing
roughly \$1.50 per subject per experiment); and second, although our panel
size restricted the current study to networks of $n=24$, in principle this
limit can be raised arbitrarily, allowing for the study of much larger
networked systems.  The speed, efficiency, and scalability of web-based
experimentation should allow researchers to extend the current study in
numerous directions: how would contributions be affected by giving
players more information about the network, or providing players with
feedback, or allowing players to rewire their network ties?  And how do
all these effects scale with the size and density of the network?  In
addressing these questions, and others, we anticipate that web-based
platforms like that provided by AMT will become an increasingly valuable
tool for understanding the dynamics of human cooperation, and for
experimental social science in general.
% 
% You may title this section "Methods" or "Models". 
% "Models" is not a valid title for PLoS ONE authors. However, PLoS ONE
% authors may use "Analysis" 

% 
\section*{Materials and Methods}
This section provides additional details on the Investment Game experiment, conducted on Amazon's Mechanical Turk (AMT).  
All participants were recruited on AMT by posting a HIT for the experiment,
entitled ``The Investment Game'', a neutral title that was accurate without disclosing the purpose of the experiment.
Before launching the experiment, we submitted to and complied with Yahoo!'s internal human subjects review process.  
A letter certifying the approval of our experiment has been filed with PLoS One. 
All data collected in the experiment could be associated only with participants' Amazon Turker ID, 
not with any personally-identifiable information; thus all players remained anonymous.

\subsection*{Ethics Statement}
Before participating, all subjects were required to read and acknowledged 
the following terms of use agreement (equivalent to an Informed Consent Statement).

\subsubsection*{The Investment Game Terms of Use}
You will be paid \$0.50 as a base rate plus more depending on your ability to play the game
If you have any questions at any time, please contact:
Siddharth Suri at Yahoo! Research, 111 W. 40th St., New York, NY, or by email at suri@yahoo-inc.com 
By clicking the ``I Agree" button below you affirm that you have read and understood the following Yahoo! Research Investment Game Terms of Use and Investment Game description and agree to comply with and be bound by its terms.
YAHOO! RESEARCH INVESTMENT GAME TERMS AND CONDITIONS
\begin{enumerate}
\item Welcome to the Yahoo! Research Investment Game ("Project"). This Project is a game of skill, not a game of chance. By participating in the Project you are entering into a legally binding agreement with Yahoo!, Inc., ("Yahoo!, ``we," ``our," and ``us"). This agreement is comprised solely of these Terms and Conditions (``Agreement" or ``Terms"), including anything explicitly incorporated by reference. If you do not agree to these Terms, please do not participate. 
\item The Project is offered to individuals registered as ``workers" with Amazon, Inc.'s ``Mechanical Turk" service (http://www.mturk.com/mturk/welcome). 
\item Your participation in the Project as a worker is governed by Amazon, Inc.'s Mechanical Turk's conditions of use (http://www.mturk.com/mturk/conditionsofuse) in addition to the following Yahoo! terms: 
\begin{enumerate}
\item Description of Project. The Yahoo! Research Investment Game is intended to collect data on how well people play this game. 
\item Work Product/Ownership. You agree to perform the tasks provided in the Project and to be compensated for the completion of each task as set forth in C below. You also agree that Yahoo!, and not You, shall own all work product from your participation in the Project. 
\item Payment. You will be paid \$0.50 plus a bonus depending on your skill level for the game. All payments will be made to You through the Mechanical Turk service as detailed in the Mechanical Turk conditions of use. 
\item Relationship of the Parties. The Parties are independent contractors. Nothing in these Terms shall be construed as creating any agency, partnership, or other form of joint enterprise between the Parties and neither Party may create any obligations or responsibilities on behalf of the other Party. 
\item Termination. 
\begin{enumerate}
\item By You. You may terminate Your participation in the Project by clicking the Return HIT button at any time. 
\item By Yahoo!. We may suspend or terminate the Project at any time, with or without notice, for any reason or no reason. In the event of such termination, Yahoo! will pay You for all tasks fully completed by You prior to termination.
\end{enumerate}
\item Contact. If you have any questions at any time, please contact Siddharth Suri at Yahoo! Research, 111 W. 40th St., New York, NY, or by email at suri@yahoo-inc.com 
\item Confidentiality. You will not disclose or use Yahoo!'s Confidential Information. ``Confidential Information" means any information disclosed or made available to You by Yahoo!, directly or indirectly, whether in writing, orally or visually, other than information that: (a) is or becomes publicly known and generally available other than through Your action or inaction or (b) was already in Your possession (as documented by written records) without confidentiality restrictions before you received it from Yahoo!. Confidential Information includes, but is not limited to, all information contained within the Project, these Terms, the Policies, and any other technical or programming information Yahoo! discloses or makes available to you. 
\item Indemnity. 
You will defend, indemnify and hold harmless Yahoo! Inc., and its affiliated companies, (``Indemnified Parties") from and against any and all claims, liabilities, losses, costs, and expenses, including reasonable attorneys' fees, which the Indemnified Parties suffer as a result of claims that arise from or relate to your activities under or in connection with this Agreement, including but not limited to claims that allege or arise from: (i) a violation a third party's right of privacy, or infringement of a third party's copyright, patent, trade secret, trademark, or other intellectual property rights, (ii) any breach of your obligations, covenants, warranties or representations as set forth in this Agreement, including any breach of any applicable policies, (iii) any violation of applicable laws, rules, and regulations by you, including, without limitation, privacy laws, and (iv) any breach of this Agreement. You shall not enter into any settlement that affects any Indemnified Party's rights or interest, admit to any fault or liability on behalf of any Indemnified Party, or incur any financial obligation on behalf of any Indemnified Party without that Indemnified Party's prior written approval. 
\item No Warranty.
YOU EXPRESSLY AGREE TO THE FOLLOWING WARRANTY DISCLAIMER. YOU ARE PARTICIPATING IN THE PROJECT AT YOUR OWN RISK. YOU REPRESENT AND WARRANT THAT BY PARTICIPATING IN THIS PROJECT THAT YOU WILL COMPLY WITH ALL APPLICABLE LAWS. THE PROJECT AND EVERYTHING PROVIDED UNDER THIS AGREEMENT IS PROVIDED ``AS IS." YAHOO! DOES NOT WARRANT THAT THE PROJECT WILL OPERATE UNINTERRUPTED OR ERROR-FREE. YAHOO AND ITS LICENSORS ARE NOT RESPONSIBLE FOR ANY CONTENT PROVIDED HEREUNDER. TO THE EXTENT ALLOWED BY LAW, YAHOO! AND ITS LICENSORS MAKE NO WARRANTY OF ANY KIND, WHETHER EXPRESS, IMPLIED, STATUTORY OR OTHERWISE, INCLUDING WITHOUT LIMITATION WARRANTIES OF MERCHANTABILITY, FITNESS FOR A PARTICULAR PURPOSE, AND NONINFRINGEMENT. YAHOO! MAKES NO WARRANTY AND NO REPRESENTATION ABOUT THE AMOUNT OF MONEY YOU WILL EARN THROUGH THE PROGRAM. THIS WARRANTY DISCLAIMER SHALL APPLY TO THE MAXIMUM EXTENT PERMITTED BY LAW. 
\item Limitation of Liability. 
YOU EXPRESSLY AGREE TO THE FOLLOWING LIMITATION OF LIABILITY. YAHOO! WILL NOT BE LIABLE FOR ANY LOST PROFITS, COSTS OF PROCUREMENT OF SUBSTITUTE GOODS OR SERVICES, OR FOR ANY OTHER INDIRECT, SPECIAL, INCIDENTAL, EXEMPLARY, PUNITIVE OR CONSEQUENTIAL DAMAGES ARISING OUT OF OR IN CONNECTION WITH THIS AGREEMENT, HOWEVER CAUSED, AND UNDER WHATEVER CAUSE OF ACTION OR THEORY OF LIABILITY BROUGHT, EVEN IF YAHOO! HAS BEEN ADVISED OF THE POSSIBILITY OF SUCH DAMAGES. YAHOO! WILL NOT BE LIABLE FOR DIRECT DAMAGES IN EXCESS OF ANY AMOUNT THAT YAHOO! HAS ALREADY PAID TO YOU FOR YOUR PARTICIPATION IN THE PROJECT. IF YOU ARE DISSATISFIED WITH ANY ASPECT OF THE PROJECT, OR WITH ANY OF THESE TERMS OF USE, YOUR SOLE AND EXCLUSIVE REMEDY IS TO DISCONTINUE YOUR PARTICIPATION IN THE PROJECT. This limitation of liability shall apply to the maximum extent permitted by law. 
\item No Public Statements. 
You may not issue any press release or other public statement regarding the Agreement, Yahoo!, and/or Yahoo! Inc.’s affiliates, or partners or advertisers without the prior written consent of an authorized person at Yahoo!.
\end{enumerate}
\end{enumerate}

I AGREE

\subsection*{Participant Instructions}

After accepting the HIT and agreeing to the terms of use, participants were provided with the following instructions (adapted from~\cite{fehr-punishment}). 

\subsubsection*{Welcome to the Investment Game!}

Because the amount of money you can earn depends on your decisions in the game, it is important that you read these instructions with care. At the end of the instructions there is a quiz to ensure that you understand the instructions. You will not be paid for the HIT unless you correctly answer these questions. 

\subsubsection*{Overview}

In the Investment Game you will be placed in a network with 23 other Turkers; however, you will only ``see" a subset of the total network-those players to whom you are connected directly. These players will be called your ``neighbors". Both the total network and your neighbors will remain fixed throughout the game.

Once the network is populated with Turkers, the game will proceed over the course of 10 ``rounds". During each round you and your neighbors (i.e. the Turkers directly connected to you in the network) will choose how much to contribute to an abstract project. Then this project generates a ``payoff" that will then be split equally among you and those who are directly connected to you. Your total payoff for the game is the sum of your payoffs from each round.

During the game we will not report your earnings in terms of dollars and cents but rather in terms of points. At the end of the game the total amount of points you have earned will be converted to dollars at the rate of 1 point = 2 cents. The amount you earn from the game will be the bonus for this HIT. You will earn the base rate of 50 cents for this HIT by correctly answering the quiz at the end of these instructions.

\subsubsection*{How the game works}

\begin{enumerate}
	\item In each round we give you an ``endowment" of 10 points. 
	\item You decide how many points you want to contribute to the project by typing a number between 0 and 10 in the input field and then clicking the submit button. Please note that by deciding how many points to contribute to the project, you also decide how many points you keep for your self, this is (10 - your contribution) points. Also note that once you have submitted your contribution you cannot go back and change it.
	\item In the first two rounds you have 45 seconds to make you contribution. In the remaining rounds you have 30 seconds. If you do not make a contribution before the end of a round, the system will make one for you and you will not earn any points for that round. 
	\item Your income from each round consists of two parts: 
	\begin{enumerate}
		\item the points which you have kept for yourself ("income from points kept"). 
		\item ``income from the project", which is 0.4 x the total contribution that you and your neighbors made to the project. 
	\end{enumerate}
	Your income in points from a round is therefore:
	Income from points kept + Income from the project = (10 - your contribution to the project) + 0.4*(total contributions you and your 	neighbors made to the project) 
\item The income of each person in the network (including your neighbors) is calculated in the same way. 
\end{enumerate}

\subsubsection*{Four Examples of Payoffs}

\begin{enumerate}
	\item Suppose you have four (4) neighbors, and each of you contributes the maximum allowable of 10 points. The sum of the contributions you and your neighbors (those who are directly connected to you) is 50 points, and so each member of the group receives an income from the project of: 0.4*50 = 20 points. Meanwhile your income from points kept = 0 (because you did not keep any), and so your total income = 0 + 20 = 20 points. 
	\item Alternatively, suppose that each player contributes two (2) points. Then the total contribution to the project is 10 points, and each member of the group receives an income from the project of: 0.4*10 = 4 points. Because you contributed two of these points then your income from points kept is eight (8), and your total income = 8 + 4 = 12 points. 
	\item Next, say you contribute two (2) points and all your neighbors contribute ten (10) points, the total contribution is 42 points, and the income that each player receives from the project is 0.4*42 = 16.8 points. Because you contributed two (2) points, your kept income is eight (8), and your total income = 8 + 16.8 = 24.8 points. 
	\item Finally, say you contribute ten (10) points, and all your neighbors contribute two (2) points, the total contribution is 18 points, and the income that each player receives from the project is 0.4*18 = 7.2 points. Because you contributed ten (10) points, your kept income is zero (0) points and your total income = 0 + 7.2 = 7.2 points. 
\end{enumerate}

\subsubsection*{Important Points to Note}

\begin{enumerate}
	\item For each point that you decide to keep for yourself, your income for that round will increase by one point. 
	\item For each point you contribute to the project, the total contribution to the project will rise by one point, and your income from the project will rise by 0.4*1 = 0.4 points. 
	\item For each point you contribute to the project, the income of your neighbors will rise by 0.4 points each. For example, if you have 4 neighbors then a one point contribution by you will raise the total income of you and your neighbors by 5*0.4 = 2.0 points. 
	\item Finally, you also earn an income for each point contributed by your neighbors to the project. For each point contributed by each of them, you earn 0.4*1 = 0.4 points. 
\end{enumerate}

\subsection*{Participant Quiz}

Finally, participants were required to pass a quiz, thus demonstrating that they had understood the instructions. 

\subsubsection*{Quiz}
To make sure you have read and understood the instructions, you must answer the following questions correctly. If you answer any questions incorrectly, you will get a second chance. If you answer a question incorrectly twice, you will not be allowed to play the game and will not receive payment for the HIT. The answers to all of the questions below are in terms of points. Please accept the HIT before beginning to fill out the form. 

In questions 1-4, assume you have 5 neighbors and you and your neighbors have an endowment of 10 points each. 
\begin{enumerate}
	\item If nobody (including yourself) contributes any points to the project what would your total income be? 
	\item If everyone (including yourself) contributes all 10 points to the project, would your total income be? 
	\item Say together your neighbors contribute a total of 25 points to the project. 
	\begin{enumerate}
		\item If you do not contribute any points to the project what would your total income be? 
		\item If you contribute an additional 5 points to the project what would your total income be? 
	\end{enumerate}
	\item Say you contribute 8 points to the project. 
	\begin{enumerate}
		\item What would be your income if your neighbors contributed a total of 12 points to the project? 
		\item What would be your income if your neighbors contributed a total of 32 points to the project? 
	\end{enumerate}
\end{enumerate}

% Do NOT remove this, even if you are not including acknowledgments
\section*{Acknowledgments}
The authors are grateful to E. Fehr for sharing the data and subject
instructions from~\cite{fehr-punishment}; and to S. Goel, D. Goldstein and
especially W. Mason for helpful conversations.  The first author would also
like to thank Claudia Neri and Peter Pal Zubcsek for helpful comments on an
earlier draft of this paper.

%\section*{References}
% The bibtex filename

\end{document}